\documentclass[conference]{IEEEtran}
\ifCLASSINFOpdf
\else
\fi
\begin{document}
%
\title{The role of self-touch experience in the formation of the self}


\author{\IEEEauthorblockN{Matej Hoffmann}
\IEEEauthorblockA{Department of Cybernetics, Faculty of Electrical Engineering, Czech Technical University in Prague\\
Email: matej.hoffmann@fel.cvut.cz\\
iCub Facility, Istituto Italiano di Tecnologia, Genova, Italy}
}

\maketitle

\begin{abstract}
The human self has many facets: there is the physical body and then there are different concepts or representations supported by processes in the brain such as the ecological, social, temporal, conceptual, and experiential self. The mechanisms of operation and formation of the self are, however, largely unknown. The basis is constituted by the ecological or sensorimotor self that deals with the configuration of the body in space and its action possibilities. This self is prereflective, prelinguistic, and initially perhaps even largely independent of visual inputs. Instead, somatosensory (tactile and proprioceptive) information both before and after birth may play a key part. In this paper, we propose that self-touch experience may be a fundamental mechanisms to bootstrap the formation of the sensorimotor self and perhaps even beyond. We will investigate this from the perspectives of phenomenology, developmental psychology, and neuroscience. In light of the evidence from fetus and infant development, we will speculate about the possible mechanisms that may drive the formation of first body representations drawing on self-touch experience.   
\end{abstract}


%
\IEEEpeerreviewmaketitle

\section{Introduction}
 


Ulric Neisser distinguishes five different selves: the ecological self, the interpersonal self, the extended self, the private self, and the conceptual self \cite{Neisser1988}. All the ``selves'' are, perhaps to different degrees, embodied and thus closely related to the representations of our bodies in our brains. Also here a number of concepts including superficial and postural schema \cite{Head1911}, body schema, body image, corporeal schema, etc. have been put forth. One characteristic common to all these representations is their multimodal nature: they dynamically integrate information from different sensory modalities (tactile, proprioceptive, vestibular, visual), not excluding motor information. Somatosensory (tactile and proprioceptive) information constitutes an important subset, perhaps most intimately tied to the body itself. However, the concepts of body schema, body image, and many others are umbrella notions for a range of observed phenomena rather than a result of identification of specific mechanisms. The field is thus in a somewhat ``chaotic state of affairs'' \cite{Berlucchi2009}. The higher level facets of the self---accessible to consciousness, incorporating linguistic and visual information---have been receiving relatively more attention. However, here we will argue that the ecological or sensorimotor self constitutes a key foundation for the rest and we will speculate about self-touch experience as a key enabler that may bootstrap its formation.  

\section{Philosophical facets: self-touch in phenomenology}
The special nature of touch---and in particular self-touch---has been appreciated by phenomenologists. In what follows we draw on the treatment of Ratcliffe \cite{Ratcliffe2013}. Both Husserl and Merleau-Ponty investigate the situation of two hands touching each other. Husserl \cite[p.~153]{Husserl1989} notes that ``the sensation is doubled in the two parts of the Body [Leib], since each is precisely for the other an external thing that is touching and acting upon it, and each is at the same time Body'' and points out the duality of perceiver and perceived in this ``double sensation''. Merleau-Ponty provides a related account in several places (e.g., \cite{Merleau1962,Merleau1968}) and emphasizes the ``reversibility'' of the roles of perceiver and perceived. Ratcliffe \cite{Ratcliffe2013} further argues against this differentiation of perceiver vs. perceived. 

Touch can perhaps be regarded as phenomenologically the most primary of senses \cite{Jonas1954,Ratcliffe2013}: Unlike vision, which possibly---at least to some extent---allows for a detached perception of the world, with touch this is not possible: touch always embraces the whole continuum between perceiving body and perceived body and ``is partly constitutive of the sense of reality and belonging, whereas other kinds of sensory experience presuppose it'' \cite{Ratcliffe2013}.
Finally, Ratcliffe \cite{Ratcliffe2013} rightly notes that ``touch'' in this context should probably not be viewed in the narrow sense of cutaneous sensation, but should also include proprioception (sense of position and movement of the body)---what is embraced by somatosensation (which we will use here to denote touch and proprioception together). 





\section{Self-touch in development: fetus and infant} \label{sec:psychology}
Not only does touch (somatosensation, more precisely) have a certain ``phenomenological primacy'', but it is also the first sense to emerge in the fetus \cite{Bradley1975}. Furthermore, it is frequently stimulated: fetuses are constantly touched by their environment (amniotic fluid, uterine surface, placenta, cord) and also engage in self-touch. 
They perform local movements directed to areas of the body most sensitive to touch: the face (the mouth in particular), but also for example soles of feet \cite[p.~ 113-114]{Piontelli2015}. Later, from 26 to 28 weeks (of gestational age), they also use the back of the hands to touch as well as touch other body areas like thighs, legs, and knees \cite[p.~ 29-30]{Piontelli2015}. In addition, from 19 weeks, fetuses anticipate the hand-to-mouth movements \cite{Myowa2006} (the mouth opens prior to contact) and from 22 weeks, the movements seem to show the recognizable form of intentional actions, with kinematic patterns that depend on the goal of the action (toward mouth vs. toward eyes) \cite{Zoia2007}.
Next to local movements toward the body and the somatosensory stimulation induced, Piontelli \cite[p. 115]{Piontelli2015} further speculates that ``turbulent general movements, with their often complete rotations (especially during the early stages of pregnancy), probably endow fetuses with a sense of their bodies in space.'' Here, vestibular inputs may complement the information from touch and proprioception.


Birth obviously brings about a major disruption of the equilibrium that was reached in the womb: the constrained aquatic environment is suddenly replaced by an aerial one, with gravity playing a major part. Nevertheless, hand-mouth coordination continues to develop after birth (e.g.,~\cite{Rochat1993}). Also, Thomas et al. \cite{Thomas2015}, biweekly recording resting alert infants from birth to 6 months of age, show that infants do frequently touch their bodies, with a rostro-caudal progression as they grow older: Head and trunk contacts are more frequent in the beginning, followed by more caudal body locations including hips, then legs, and eventually the feet. Rochat \cite{Rochat2001,Rochat2011} also discusses a ``two-month revolution'': the emergence of a more contemplative stance taken by infants in their attention to events and things, correlated with a sudden increase in the time the infant spends awake and alert \cite{Wolff1987}.

In order to gain some insight into how much ``functional body knowledge'' the infants have at different ages, Lockman et al. invented an experimental paradigm whereby vibrating stimuli (buzzers) are attached to different body parts and the infant's responses---e.g., movement of a stimulated body part or reaching action toward the buzzer---are observed \cite{Leed2014,Hoffmann2017icdl,Somogyi2017}. The results indicate that there is dramatic development in particular in the period between 3 and 8 months of age.

\section{Self-touch and the brain} \label{sec:brain}
The development of brain circuits is obviously another key determinant of the construction of the ``self'' in the fetus, infant, and child. Tau and Peterson \cite{Tau2010} provide an excellent survey of the processes and their timing in this period (neural proliferation and migration, apoptosis, synaptogenesis, and myelination). Relating specifically to the processing of somatosensation, around 18-22 weeks, neurons receiving preliminary afferent inputs from the somatosensory thalamus are located on the subplate (a transient embryonic cortical layer). Synaptic connections here serve as placeholders ahead of later connections for thalamocortical neurons \cite{Tau2010} (for more details on these see \cite{Kostovic2010}). Milh et al.~\cite{Milh2007} show in premature neonates that spontaneous movements give rise to stimulation in the somatosensory cortex in a somatotopic manner. 


The early period after birth represents a time of dramatic change in brain structure and function. Next to massive overall growth of brain volume, the perinatal and postnatal brain development features also the onset of myelination and striking development (but also pruning) of gray matter connections, especially in the sensorimotor and visual cortices \cite{Tau2010}. Experience (or activity-dependent factors) starts playing a more prominent function in the shaping of neural circuitry \cite{Khazipov2006,Tau2010}. While primary cortical areas (including motor, somatosensory, visual, and auditory cortices) can be identified, association cortices are less clearly delineated in the newborn brain \cite{Tau2010}.

Around 2 months of age, there is a transition from the dominant control of behaviors by subcortical systems to higher order cortical systems \cite{McGraw1943}.
PET imaging studies suggest that subcortical regions and the sensorimotor cortex are the most metabolically active brain regions in neonates younger than 5 weeks; by 3 months of age, metabolic activity then increases in parietal, temporal, and dorsolateral occipital cortices \cite{Chugani1987,Chugani1994}. These facts may be in line with (or even bring about) the ``two-month revolution'' in infant behavior reported above. 



All the above mentioned facts suggest that already the prenatal brain should be able to perceive stimulations arising from events pertaining to the body and its configuration. Higher-level or multimodal representations may possibly form somewhat later in the course of the first year after birth, as these are thought to reside in association cortices (in particular posterior parietal areas). It is unclear whether the self-touch events play a special part in the ``learning of a body model''. Interestingly, albeit in adult monkey brain, Sakata et al. \cite{Sakata1973} found neurons in area 5 of the parietal cortex that responded specifically to self-touch: ``'matching neurons', because the 'best' stimulus for them was to bring a part of a limb in contact with a part of another limb or the trunk, as if to match two body regions to each other.'' The response was elicited only in specific body configurations (as perceived through proprioceptive afference) and only if the corresponding tactile stimulation was present. 

\section{Mechanisms of learning a body model from self-touch}
Perhaps the redundant information induced by the self-touch configurations in the proprioceptive-tactile space (or motor-proprioceptive-tactile-visual if the whole system is considered and the body part is visible) may facilitate learning about the body in space. Rochat \cite{Rochat1998} writes: ``By 2-3 months, infants engage in exploration of their own body as it moves and acts in the environment. They babble and touch their own body, attracted and actively involved in investigating the rich intermodal redundancies, temporal contingencies, and spatial congruence of self-perception.''  In Piaget's view, in this period (from birth to about 9 months), ``as yet, no constant relation exists between visual and buccal space or between tactile and visual space. True, auditory and visual space are already coordinated, as are buccal and tactile space, but no total and abstract space encompasses all the others.'' Then, Piaget proposes prehension (grasping) as a key enabler to connect visual space to tactile and gustatory space \cite{Piaget1954}. 

Neither the mechanisms of this developmental process nor the operation of the body models are clear. Below we first discuss the relationship between reaching to external (visual) targets and reaching to own body. Then we review the mathematical treatment of inferring the dimensionality of space (or body in space) from sensorimotor contingencies. Finally, we review robotic models thereof.

\subsection{Reaching to visual targets vs. reaching to the body}
Goal-directed reaching begins to develop only slightly later than multisensory processing, at roughly 5 months of age. While a full account is out of the scope of this paper, we want to point out the implications on reaching to one's own body. As reviewed in \cite{Williams2016} for example, researchers initially concentrated on the role of vision in guiding the arm toward a target object. Since the late 1990s they have started investigating reaching as the product of multiple interacting subsystems, with important parts played by the motor system, proprioception, and the whole embodiment and dynamics of the infant in general (e.g., \cite{Corbetta2014,Thelen1994}). Localizing targets on the own body is special in two respects: (i) the body that is used to act on the environment itself becomes a target; (ii) for certain targets, the process bypasses vision. More precisely, it has to do so in the case of non-visible targets like on the face; and it may do so with other targets, relying on somatosensory information only. While both types of encoding---visual (sometimes called extrinsic) as well as postural (intrinsic)---were found in different regions in the parietal cortex (e.g.,~\cite{Pellijeff2006}), the developmental trajectory is unclear. In particular, does reaching for somatosensory targets on the body develop first and separately from reaching to visual targets in external peripersonal space? How do these systems interact and what do they share? Heed et al. \cite{Heed2015} provide a survey of possible architectures that could support the remapping needed for action towards a tactile stimulus (from skin-based to external reference frame). In adults, coordinates drawing on vision (eye-centered or gaze-centered) seem to be involved even when processing targets on the body \cite{Harrar2010,Mueller2014}. However, considering that vision continues to develop in early infancy, it is also possible to think that localization of the body in space may first be established relying on somatosensory inputs, and this representation may later be shaped or overridden by a vision-based encoding. 

\subsection{Inferring the body in 3D space from self-touch}
One particular theoretical problem is extracting the dimensionality of external space (which is 3-dimensional) without prior knowledge from high-dimensional multimodal sensory or sensory-motor data. Work on this topic goes back to Poincar{\'e} \cite{Poincare1905} who showed that an arm with $n$ degrees of freedom equipped with encoders and with the end-effector fixed in space can through movement infer the dimensionality of external space from the real dimensionality of the proprioceptive signal spaces (as it moves in the joint space, keeping the end-effector fixed). This was followed up more recently by considering active agents collecting sensorimotor experience \cite{Laflaquiere2015,Philipona2003,Terekhov2016}. Roschin et al.~\cite{Roschin2011} proposed a model specifically addressing a self-touch scenario, fusing tactile and proprioceptive information. 

\subsection{Embodied brain models and self-touching robots}
The models reviewed in the previous section have a theoretical nature and speak mostly to the question of whether extracting 3D space and a spatial representation of the body in this space is theoretically possible. However, they do not imply that these mechanisms are used by biological organisms and their neural systems. To this end, embodied brain models are necessary. Yamada, Kuniyoshi et al. have addressed fetal development, including touch and the formation of somatosensory representations, in a series of fetal simulators coupled with brain models, culminating in \cite{Yamada2016}. We have pursued a similar approach targeting early postnatal development in the iCub humanoid robot: representations of the tactile space \cite{HoffmannStraka2017} and a preliminary attempt toward a proprioceptive representation \cite{HoffmannBednarova2016} were learned in a bottom-up fashion using self-organizing maps. The next steps will involve connecting these modalities using self-touch (already implemented in the robot relying thus far on engineered modules \cite{Roncone_ICRA_2014}).



\section{Discussion and Conclusion}

The treatment of self-touch by phenomenologists suggests a special role that these events may play in the formation of the different types of the self, including the experiential self. In the rest of the paper, we focused on the sensorimotor self only though. The redundant information induced by self-touch configurations in the proprioceptive-tactile space (or motor-proprioceptive-tactile-visual if the whole system is considered and the body part is visible) should suffice to facilitate learning about the body in space. Furthermore, as the self-touch (or double-touch) configurations are unique---with tactile stimulation on two different body parts and only in specific joint configurations---they might constitute a ``contingent stimulus'', associated with a reward or neuromodulation that bootstraps learning. Observations of fetal and infant behavior (Section \ref{sec:psychology}) evidence frequent spontaneous self-touch events both before and after birth. In Section~\ref{sec:brain}, we reviewed some of the facts about the readiness of the sensory and nervous system to process self-touch at different periods of development. Putting this together, we speculate that self-touch experience in the womb probably contributes to the formation of primary somatosensory representations. The anticipation of hand-to-mouth movements indicates also first instances of body models that span multiple modalities. However, the formation of more holistic multimodal representations of the body in space occurs probably only after birth, in particular from about 2-3 months. The evidence supporting this claim includes: (i) the perinatal period involves dramatic changes to the body itself, the environment it is in, and brain development, which makes a ``transfer'' of a body model unlikely; (ii) the representations of the body in space are believed to reside in the association areas of posterior parietal cortex, which are developing mostly after birth and later than unimodal areas; (iii) cortical control of movement is also taking over from around 2 months after birth. This hypothesis is in line with our observations when attaching buzzers to infant body parts, whereby there is dramatic development in particular in the period between 3 and 8 months of age \cite{Leed2014,Hoffmann2017icdl,Somogyi2017}. Finally, computational models corroborate the possibility that a model of the body in space can be extracted from somatosensory (or motor-somatosensory) information only. Embodied brain and robotic models make it possible to further ground the computational models and put the hypotheses proposed here to test. 








\section*{Acknowledgment}
This work was supported by a Marie Curie Intra European Fellowship (iCub Body Schema 625727) within the 7th European Community Framework Programme and the Czech Science Foundation under Project GA17-15697Y. I would like to thank Jacqueline Fagard for materials on fetus development, Lisa Chinn for comments on the manuscript, and Shaun Gallagher for pointers to the treatment of self-touch by Husserl, Merlau-Ponty, and Ratcliffe.



\bibliographystyle{IEEEtran}
\bibliography{Hoffmann_SelfTouchForSMself}
%



\end{document}